\documentclass[prl, aps, 10pt, showpacs, superscriptaddress, twocolumn, floatfix]{revtex4-2}
\usepackage{graphicx}
\usepackage[usenames,dvipsnames]{color}
\usepackage{siunitx}
\usepackage{enumerate}
\usepackage{amsmath}
\usepackage{amssymb}
\usepackage{hyperref}
\usepackage{braket}
\usepackage{ulem}
\usepackage{cleveref}
\usepackage{cancel}
\usepackage{soul} 
\usepackage{multirow}

\newcommand{\An}{$A_{n}$ }
\newcommand{\Pb}{$^{208}$Pb }

\newcommand{\cerenkov}{\v{C}erenkov }

\begin{document}

\title{Beam-Normal Single-Spin Asymmetry in \Pb at low energy: discrepancy resolved or new kinematic puzzle?}

\def\kph{\affiliation{Institut f\"ur Kernphysik, Johannes
  Gutenberg-Universit\"at Mainz, D-55099 Mainz, Germany}}
\def\him{\affiliation{Helmholtz Institute Mainz, GSI Helmholtzzentrum f\"ur  Schwerionenforschung, Darmstadt, Johannes Gutenberg-Universit\"at, D-55099 Mainz, Germany}}
\def\prisma{\affiliation{PRISMA$^+$ Cluster of Excellence, Johannes Gutenberg-Universit\"at, D-55099 Mainz, Germany}}
\def\italy{\affiliation{Dipartimento di Fisica "E. Fermi", Universit\'a di Pisa, Italy}}
\def\stefan{\affiliation{Jo\v zef Stefan Institute, 
    SI-1000 Ljubljana, Slovenia}}
\def\Ljub{\affiliation{Faculty of Mathematics and Physics, University of Ljubljana, SI-1000, Ljubljana, Slovenia}}

\author{A.~Esser}\kph
\author{N.~Kozyrev}\kph
\author{K.~Aulenbacher}\kph \him \prisma
\author{S.~Baunack}\kph
\author{M.~Dehn}\kph
\author{A.~Del Vincio} \kph \italy
\author{L.~Doria}\kph
\author{M.~Hoek}\kph
\author{F.~Keil}\kph
\author{F.~Maas}\kph \him \prisma
\author{H.~Merkel}\kph \prisma
\author{M.~Mihovilovi\v{c}}\stefan \Ljub
\author{U.~Müller}\kph
\author{J.~Pochodzalla}\kph \him
\author{B.~S.~Schlimme}\kph
\author{T.~Shao}\kph
\author{S.~Stengel}\kph
\author{M.~Thiel}\kph
\author{L.~Wilhelm}\kph
\author{C.~Sfienti}\kph \him \prisma

\begin{abstract}
A longstanding discrepancy between measured and predicted beam-normal single-spin asymmetries \An in elastic electron scattering off \Pb has challenged our understanding of two-photon exchange (TPE) in heavy nuclei. We report a new measurement at \qty{570}{\MeV} and \(Q^2 = \qty{0.04}{\GeV^2}/c^2\), yielding $A_n = (-9.1 \pm 2.1~\text{(stat)} \pm 0.7~\text{(syst)})~\mathrm{ppm}$.
This nonzero value contrasts with previous results at higher energies and suggests a kinematic dependence of TPE effects not captured by current theory, prompting a reevaluation of earlier interpretations.
\end{abstract}


\maketitle


Understanding the fundamental dynamics of electron scattering is essential for precision tests of the Standard Model, the determination of nucleon and nuclear structure, and the interpretation of a wide range of experimental observables. While the leading-order one-photon exchange approximation provides a successful description of most processes, higher-order contributions — particularly those arising from two-photon exchange (TPE) — have emerged as a critical source of theoretical uncertainty~\cite{Marciano:1980,Carlson:2007}.
TPE processes, in which the electron interacts via the exchange of two virtual photons with the target, encode essential information about hadronic structure and dynamics that is not accessible at leading order. They have been invoked to explain longstanding discrepancies in the extraction of proton electromagnetic form factors~\cite{Blunden:2003}, contribute to theoretical uncertainties in parity-violating electron scattering (PVES) - where TPE enters alongside dominant $\gamma$–$Z$ box corrections~\cite{Gorchtein:2009}- , and crucially enter the interpretation of muonic atom spectroscopy~\cite{Antognini:2013review} — where they currently limit the precision with which nuclear charge radii can be determined. Despite their relatively small size, typically at the percent level, TPE effects become decisive whenever experiments push the limits of precision.
A rare opportunity to isolate TPE effects is offered by the beam-normal single-spin asymmetry, $A_n$.
This asymmetry arises in elastic electron scattering when the incident beam is polarized normal to the scattering plane (often referred to as ‘beam-normal polarization’). It is defined as
\begin{equation}
A_n = \frac{\sigma^\uparrow - \sigma^\downarrow}{\sigma^\uparrow + \sigma^\downarrow},
\end{equation}
where $\sigma^\uparrow$ and $\sigma^\downarrow$ are the cross sections for spin parallel and antiparallel to the normal vector $\hat{n} = (\vec{k} \times \vec{k}^\prime)/|\vec{k} \times \vec{k}^\prime|$ of the scattering plane, with $\vec{k}$ and $\vec{k}^\prime$ the three-momenta of the incoming and scattered electrons, respectively.

At leading order \An arises from the imaginary (absorptive) part of the interference between the one- and two-photon exchange amplitudes~\cite{DERUJULA1971365}.
Several theoretical approaches have been developed to calculate $A_n$ in electron scattering. 
For the $p(e,e’)p$ reaction, multiple methods are available~\cite{Pasquini:2005, Afanasev:2004}. However, for nuclear targets with $Z \geq 2$, only two established models exist. The Coulomb-distorted wave approach~\cite{Cooper:2005} includes all orders of photon exchange but only the nuclear ground state. The dispersion relation framework~\cite{Gorchtein:2008} incorporates excited intermediate states but is valid only at low momentum transfer and introduces uncertainty through the poorly constrained Compton form factor.
In both cases, predictions for heavy nuclei rely on extrapolations and assumptions that may not hold across all regimes.
This limitation became evident with the unexpected result from the PREX collaboration~\cite{Abrahamyan:2012}, which reported a vanishing \An in $^{208}$Pb at $E = 1.063$ GeV and $Q^2 \sim 0.009~\mathrm{GeV}^2/c^2$, in clear disagreement with theoretical expectations. The discrepancy was later confirmed at similar kinematics by PREX-II/CREX~\cite{Adhikari:2022}, but remains unexplained.

Motivated by this anomaly, a measurement program using the A1 high-resolution spectrometers~\cite{Blomqvist:1998xn} at MAMI~\cite{ankowiak2006-ou} was launched to study $A_n$ across different nuclear targets~\cite{Esser:2018,Esser:2020}. Results for $^{12}$C, $^{28}$Si, and $^{90}$Zr were generally consistent with the dispersion-based prediction within its estimated 20\% uncertainty. No dramatic suppression of \An—like that seen in $^{208}$Pb—was observed. 
In this Letter, we present the first measurement of \An in elastic electron scattering from $^{208}$Pb at $E = 570$ MeV and $Q^2 = 0.04~\mathrm{GeV}^2/c^2$. The kinematics match earlier measurements on lighter nuclei, enabling a direct comparison and testing the nuclear and kinematic dependence of TPE.


To measure \An a vertically polarized continuous-wave electron beam with an energy of \qty{570}{\MeV} was scattered from a \Pb foil of \qty{0.5}{\mm} thickness (\qty{567}{\mg\per\square\cm}), corresponding to about one-tenth of a radiation length. This thickness optimized the reaction rate while minimizing depolarization from multiple scattering and target damage. The foil was mounted in a copper frame connected to a heat exchanger cooled to 5$^\circ$C. Fast steering magnets rastered the beam over an area of approximately \qtyproduct{4 x 4}{\milli\metre} to reduce local heating; the raster signals were synchronized with the readout gates to suppress false asymmetries from thickness variations.

Two magnetic spectrometers (A and B) were positioned symmetrically around the beam pipe at forward angles, defining a momentum transfer of \(Q^2 = \qty{0.04}{\GeV^2}/c^2\), matching earlier measurements on lighter nuclei. At this $Q^2$, the elastic cross section for \Pb shows a local minimum, resulting in lower rates than in previous measurements on $^{12}\mathrm{C}$ and $^{28}\mathrm{Si}$~\cite{Esser:2018,Esser:2020}. Elastically scattered electrons were detected in fused-silica \cerenkov detectors placed in the focal planes of the high-resolution spectrometers, read out by UV-sensitive PMTs. Each detector consisted of multiple PMTs optically coupled to the radiator bars. Initial alignment and calibration data were taken at low current (\qty{50}{\nano\ampere}) with tracking detectors, which were then switched off during physics runs at \qty{20}{\micro\ampere} to prevent damage.

A key component of this experiment was the newly developed data acquisition system, described in detail in Ref.~\cite{Esser:2025}. Specifically designed for high-precision asymmetry measurements at low rates, it employed four modular FPGA boards to count individual PMT pulses instead of integrating a current, thereby reducing statistical uncertainty. A master FPGA synchronized readout gates to the 50 Hz power grid frequency and generated pseudo-random polarity sequences using de Bruijn patterns~\cite{debruijn} to suppress systematic effects. PMT signals were processed using NINO discriminators~\cite{ANGHINOLFI2004183} with channel-specific attenuation, and threshold settings which were optimized through dedicated scans based on pulse-height spectra.

Beam diagnostics were integrated via voltage-to-frequency converters located close to the monitors, enabling high-resolution digitization of current, position, and energy. This minimized analog noise and allowed event-by-event asymmetry corrections. 

\begin{figure}[h]
\centering
\includegraphics[width=0.48\textwidth]{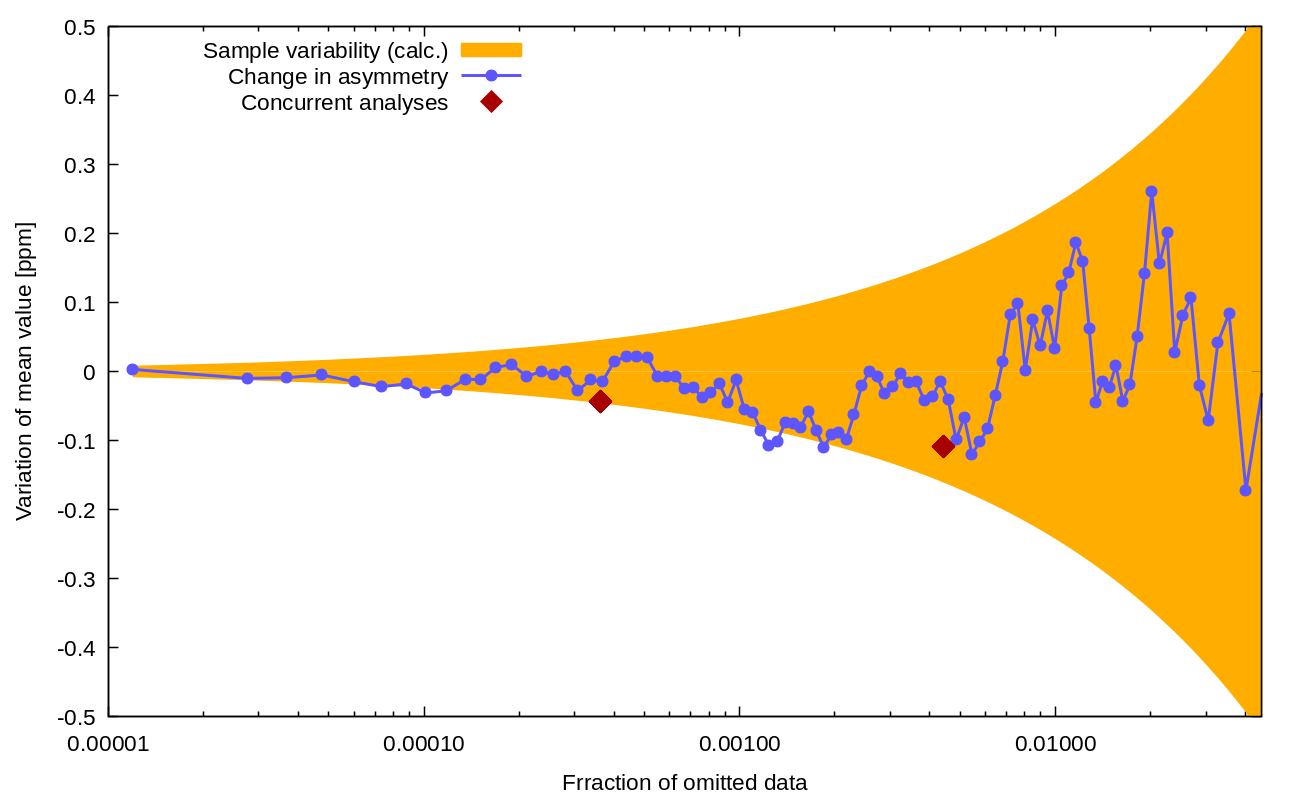} 
\caption{Change of the measured asymmetry versus the fraction of dismissed data, for spectrometer A.
The orange trumpet shape shows the expected sample variability of the mean.
Blue points represent a sequence of cuts of increasing strictness.
The two red diamonds indicate the values obtained by the two analysis chains.}
\label{fig:trumpetA}
\end{figure}

Beam polarization was measured at the start of each experimental campaign using a Møller polarimeter, cross-calibrated against a Mott system at low energy~\cite{SCHLIMME201754}, providing continuous tracking of the degree and orientation of polarization. 
Two separated experimental campaigns (November 2023 and June 2024) allowed for stringent cross-checks of systematics; in total, 232.7 hours of data were collected on the \Pb target.
The analysis was conducted using a systematic multi-step procedure. It begins by excluding short periods in which the accelerator or beam stabilization systems were not functioning properly—such as during sudden beam losses or magnetic steering failures. This manual selection was deliberately conservative to preserve as much of the data set as possible; overall only 2.8\,\% of the events were rejected at this stage.
Next, the beam monitor signals were calibrated to relate the raw electronic outputs to physical quantities. For the beam position and energy monitors, absolute calibration is not strictly necessary for the extraction of the asymmetry, as the units cancel out in the regression-based correction. Nonetheless, calibrated values are important to quantify the scale of beam fluctuations and provide physical context. To calibrate the position monitors, the electron beam was slowly rastered across a target consisting of three carbon rods with known spatial coordinates. The position of the rods was inferred from the \cerenkov detector response, allowing the beam monitor scale to be determined separately in both horizontal and vertical directions. The beam energy monitor, which already has a known transfer function, was verified by superimposing a defined voltage offset corresponding to a known energy shift. In contrast to the position and energy monitors, offsets in the beam current monitors and in the photomultiplier count rates directly impact the extracted asymmetry and therefore required precise determination. For this purpose, an automated calibration run was executed every three hours. During these runs, the beam current was ramped up in discrete steps from \qty{17.5}{\uA} to \qty{21.5}{\uA}, while all detector signals were recorded. The current monitor signals were linearly fitted against the reference from a calibrated fluxgate current probe, and the PMT count offsets were extracted from linear fits of count rate versus beam current.\\
Following the calibrations, individual events were subjected to quality cuts. While long periods of instability had already been excluded manually, event-by-event fluctuations required an automated approach. A set of cut rules was defined to reject events with anomalous beam parameters or detector responses. 
The robustness of the final asymmetry against the strength of these cuts was tested by applying a range of thresholds, as illustrated in Fig.,\ref{fig:trumpetA}. The orange trumpet shape 
%
%
indicates the expected statistical spread of the mean under different cut strengths. To further ensure the robustness of the final analysis, the data were independently processed using two separate analysis chains, yielding consistent results (red diamonds).\\
After data selection, the experimental asymmetry $A_\mathrm{exp}$ was computed for each event. This asymmetry includes contributions from helicity-correlated beam fluctuations and must be corrected to extract the physical beam-normal single-spin asymmetry. The correction was implemented via a multi-parameter linear regression of the form:

\begin{align*}
A_n = \frac{1}{P_\perp}\cdot \Bigl( A_{exp} &- c_I A_I - c_X \Delta X - c_Y \Delta Y \\
&- c_{X^\prime} \Delta X^\prime - c_{Y^\prime} \Delta Y^\prime - c_E \Delta E \Bigr),
\end{align*}

where $P_\perp$ is the degree of vertical beam polarization, and the coefficients $c_i$ represent the sensitivity of the asymmetry to helicity-correlated differences in beam current $A_I$, horizontal and vertical position $\Delta X$, $\Delta Y$, angle $\Delta X^\prime$, $\Delta Y^\prime,$ and energy $\Delta E$. The coefficients were obtained from a simultaneous fit to the full data set, minimizing correlations between $A_\mathrm{exp}$ and the various beam parameters. In this way, all instrumental asymmetries were removed event-by-event before applying the polarization normalization.

\begin{figure}
\centering
\includegraphics[width=0.45\textwidth]{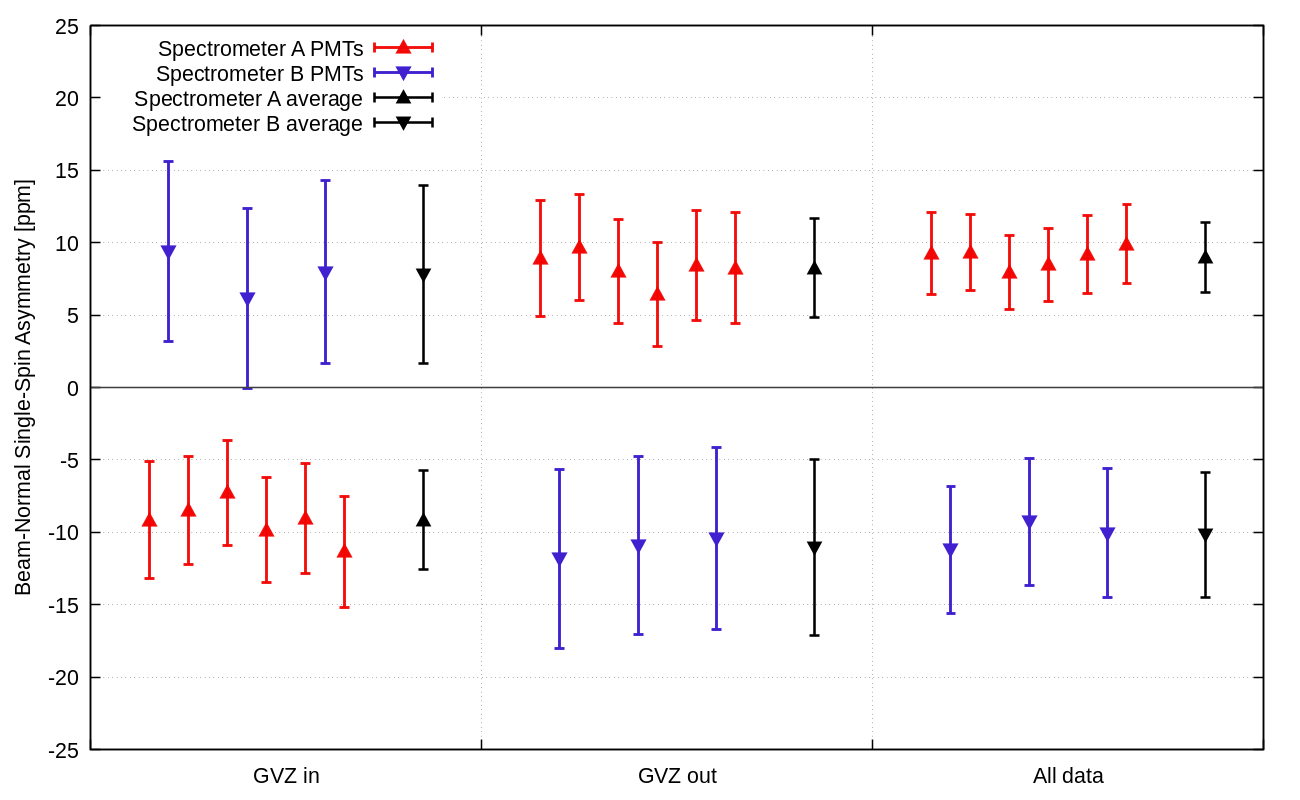}
\caption{Measured beam-normal single-spin asymmetry for both signs of the general polarity convention (GVZ in and GVZ out), and for the full data set. The general sign was reversed by inserting an additional $\lambda$/2 wave plate into the laser beam of the polarized electron source. Colored points represent individual PMTs; the black markers denote the weighted average per spectrometer. The asymmetry for spectrometer A for the full data set has been inverted.}
\label{fig:result}
\end{figure}

Finally, to combine the asymmetries from different photomultiplier channels, a weighted average was computed for each spectrometer. The weighting was based on the total number of counts recorded by each channel, ensuring that channels with higher statistical power contributed proportionally more to the final value. Due to the optical coupling of the PMTs and the adjusted discriminator thresholds, the channel-to-channel variations in weight were modest. In spectrometer A, the weights varied between 0.856 and 1.105, while in spectrometer B the variation was even smaller, ranging from 0.996 to 1.003. This reflects differences in detector geometry: spectrometer B focuses elastically scattered electrons onto a compact spot, whereas spectrometer A focuses onto an extended line, leading to a slightly broader distribution of channel contributions. 
The final asymmetries are shown in Fig.~\ref{fig:result}, including individual PMT asymmetries and the weighted averages for each spectrometer. The combination of the two spectrometers was performed using a standard error-weighted average.

\begin{table}
\centering
\caption{\label{tab:results} Measured beam-normal single-spin asymmetries for each spectrometer. The difference in scattering angle at similar $Q^2$ arises from the broader angular acceptance of spectrometer A. Uncertainties are in parts per million (ppm), with statistical and systematic contributions listed separately. 
The first five entries correspond to asymmetry correction errors. Further contributions: 
$\Delta A_{\mathrm{I}}$ estimates the residual beam-current asymmetry,  
$\Delta$Gain assesses PMT gain variations,  
$\Delta$Tails estimates for nonlinearities from large corrections,  
$\Delta$Inversion accounts for the different number of events in both states of the half-wave plate, 
$\Delta$Analysis quantifies the spread between the two independent analysis chains and
$\Delta P$ gives the polarization uncertainty.
} \vspace{3mm}
\begin{tabular}{l|c|c}
\hline \hline
Spectrometer          & A & B\\
Scattering angle      & 20.14 $^{\circ}$ & 20.56$^{\circ}$\\
$Q^2$ (GeV$^2$/$c^2$) & 0.040 & 0.041  \\
$A_{\mathrm{n}}$ (ppm)                & 8.954  & 9.568 \\ \hline  \hline
$\Delta (\partial \sigma/\partial x)$ & $< 0.001$ & $< 0.001$ \\
$\Delta (\partial \sigma/\partial y)$ & 0.012 & 0.009 \\
$\Delta (\partial \sigma/\partial x^\prime)$& $< 0.001$ & 0.006 \\
$\Delta (\partial \sigma/\partial y^\prime)$&  0.012 & 0.003 \\
$\Delta (\partial \sigma/\partial E)$       & 0.031 & 0.018 \\
$\Delta A_{\mathrm{I}}$        & 0.008 & 0.008 \\
$\Delta$Gain      & 0.034 & 0.016 \\
$\Delta$Tails      & 0.145 & 0.034 \\
$\Delta$Inversion & 0.026 & 0.074 \\ 
$\Delta$Analysis   & 0.029 & 0.611\\
$\Delta P$    & 0.109 & 0.117 \\
\hline \hline
{Total systematic error} & 0.192 & 0.628 \\ \hline \hline
Statistical error                       & 2.416 & 4.331  \\
\end{tabular}
\end{table}


%
The results for \An obtained as the arithmetic mean of the two independent analysis chains together with their uncertainties are shown in Table~\ref{tab:results}. 
The systematic uncertainty was evaluated by varying key analysis parameters and quantifying their effect on the extracted asymmetry, following the same strategy as in our previous measurements on lighter nuclei~\cite{Esser:2018,Esser:2020}. The contributions from fluctuations of the beam position, angle, and energy were estimated by varying the respective correction factors independently by $\pm$25\%, and the resulting change in \An was taken as a conservative estimate of sensitivity to imperfect beam corrections.

The contribution from residual beam-current asymmetry ($\Delta A_{\mathrm{I}}$) was estimated by adding the statistical error of its mean to the systematic budget. Fluctuations of the PMT signal offsets ($\Delta\mathrm{Gain}$) were assessed using all calibration runs: the offsets were varied within one standard deviation, and the accumulated effect was divided by $\sqrt{N_{\mathrm{calib}}}$ to account for statistical fluctuations. Possible nonlinearities in the asymmetry correction ($\Delta\mathrm{Tails}$) were estimated by excluding the 0.1\% of events with the largest absolute corrections for each term in the asymmetry equation, summing the resulting shifts in the mean asymmetry.

To test for possible instrumental asymmetries from the polarity control system, a half-wave plate in the optical system at the beam source~\cite{aulenbacher:2007} was used to reverse the beam polarization independently of the electronics. A similar number of events was collected for both half-wave plate states (GVZ in/GVZ out in Fig.~\ref{fig:result}). Although the observed difference in \An was not statistically significant, the estimated change for an equal number of events in both states was conservatively added as an additional uncertainty ($\Delta\mathrm{Inversion}$). The small spread between the almost identical results of the two independent analysis chains was conservatively included as $\Delta\mathrm{Analysis}$. The uncertainty of the beam polarization ($\Delta$P) was added separately. All contributions were summed in quadrature to obtain the total systematic uncertainty.


\begin{figure}
\centering
\includegraphics[width=0.45\textwidth]{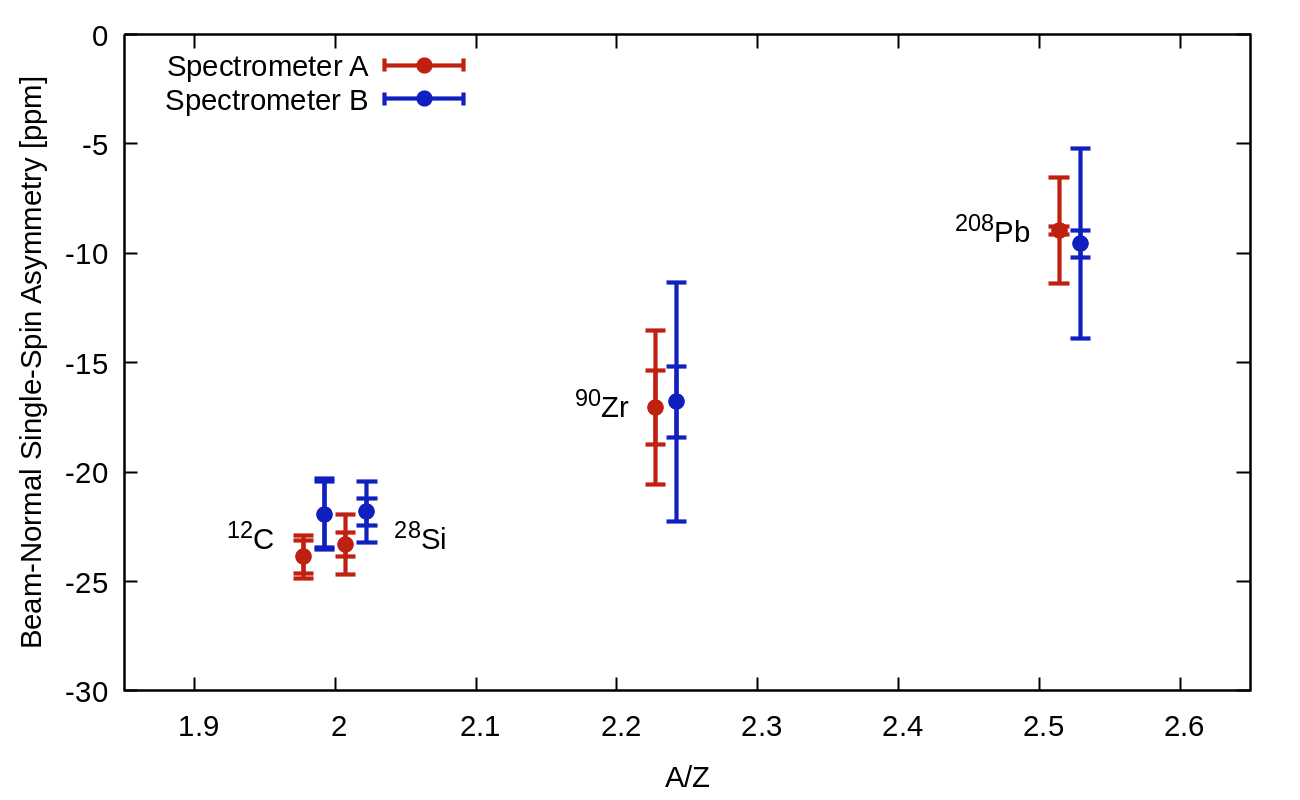}
\caption{Comparison of the beam-normal single-spin asymmetries of all nuclei measured at MAMI with similar kinematics plotted versus the atomic number over nuclear charge ratio.
For better visibility, data points with identical ratios are shifted apart.
The data points on the bottom left correspond to $^{12}$C and $^{28}$Si, the central point represents $^{90}$Zr.
The new data points for \Pb are shown on the top right.}
\label{fig:nucleiComparison}
\end{figure}

%
Our new measurement yields a beam-normal single-spin asymmetry of
\An = \text{{$-9.1 \pm 2.1_{\text{stat}} \pm 0.7_{\text{syst}}$}}~ppm
at a beam energy of \qty{570}{\MeV} and $Q^2 = 0.04~\mathrm{GeV}^2/c^2$.
 This value is significantly different from zero and contrasts with the vanishing asymmetry reported by PREX at \qty{1.06}{\GeV}.
 This suggests a non negligible energy dependence of TPE effects not captured by current theoretical models.
A comparison across all measured nuclei at MAMI shows a decrease in $\left| A_n \right|$ with the mass-to-charge ratio $\mathrm{A/Z}$, as visualized in Fig.~\ref{fig:nucleiComparison}, which may reflect indirect nuclear structure effects, such as neutron skins or inelastic contributions. 
This apparent $\mathrm{A/Z}$ scaling is kinematics-dependent: at higher beam energies, as in the CREX measurement~\cite{Adhikari:2022}, $\left| A_n \right|$ remains nearly constant up to $\mathrm{A/Z} \approx \mathrm{2.4}$ and then drops sharply for $^{208}\mathrm{Pb}$, indicating that no single parametrization can simultaneously describe both kinematic regimes.

Updated calculations using new Compton form factors~\cite{PhysRevC.103.064316} fail to reproduce either the low-energy MAMI data or the high-energy PREX result for $^{208}\mathrm{Pb}$, suggesting that high-energy input alone is insufficient to describe the observed asymmetries. 
Simplified estimates~\cite{Mishaprivate} based on the same phenomenological Compton inputs but retaining only the leading logarithmic term, predict an approximate scaling with $\mathrm{A/Z}$. However these calculations omit Coulomb distortions and subleading contributions and fail to describe the measured magnitude.

Ongoing efforts, including a proposal at the Thomas Jefferson National Accelerator Facility~\cite{JLab:PAC52}, aim to systematically test the nuclear dependence of TPE-related radiative corrections.
These discrepancies and open questions reinforce the importance of a systematic study of \An across a range of beam energies and nuclear targets. In particular, for the planned Mainz Radius EXperiment (MREX),  which aims to determine the neutron skin of \Pb at \qty{155}{\MeV} using parity-violating electron scattering, a reliable understanding of \An is essential, as it may become a dominant source of systematic uncertainty. To address this, a new measurement campaign on lead is planned at MAMI, and future opportunities at the MESA accelerator at lower energies will further constrain the role of TPE in heavy nuclei.

\begin{acknowledgments}
\noindent
We acknowledge  support from the technical staff at the Mainz Microtron and thank the accelerator group for the excellent beam quality.
This work was supported by the DFG – Project number: 514321794 (CRC1660: Hadrons and Nuclei as Discovery Tools), the DFG individual grant No. 454637981, and the Federal State of Rhineland-Palatinate.
\end{acknowledgments}

\bibliographystyle{apsrev4-1} 
\bibliography{transverse}

\end{document}